\documentclass[final,3p,times]{elsarticle}

\usepackage[utf8]{inputenc} 
\usepackage[T1]{fontenc}
\usepackage{amssymb}
\usepackage{amsmath}
\usepackage{mathrsfs}
\usepackage{graphicx,hyperref}

\renewcommand{\vec}{\boldsymbol}
\newcommand{\beq}{\begin{equation}}
\newcommand{\eeq}{\end{equation}}
\newcommand{\bea}{\begin{eqnarray}}
\newcommand{\eea}{\end{eqnarray}}
\newcommand{\baa}{\begin{array}}
\newcommand{\eaa}{\end{array}}

\def\eq#1{{Eq.~(\ref{#1})}}

\def\fig#1{{Fig.~\ref{#1}}}
\newcommand{\intl}{\int\limits}

\newcommand{\bas}{\bar{\alpha}_S}
\newcommand{\as}{\alpha_S}

\newcommand{\nn}{\nonumber}

\newcommand{\x}{\vec{x}}
\newcommand{\vb}{\vec{b}}

\newcommand{\Lb}{\left(}
\newcommand{\Rb}{\right)}

\renewcommand{\vec}[1]{\boldsymbol{#1}}

\newcommand{\dY}{\delta \tilde{Y}}

\numberwithin{equation}{section}

\begin{document}

\begin{frontmatter}



\title{Multiplicity distributions in DIS for heavy nucleus}

\author[1]{Carlos Contreras}
\ead{carlos.contreras@usm.cl}
\author[1,2]{Jos\'e Garrido\corref{speaker}}
\ead{jose.garridom@sansano.usm.cl}

\affiliation[1]{
organization={Department of Physics, Universidad T\'ecnica Federico Santa Mar\'ia},
addressline={Avenida Espa\~na 1680},
postcode={Casilla 110-V},
country={Valpara\'iso, Chile}
}

\affiliation[2]{organization={Institute of Physics, Pontificia Universidad Cat\'olica de Valpara\'iso},
addressline={Avenida Universidad 330},
country={Curauma, Valpara\'iso, Chile}}

\date{\today} 
\cortext[speaker]{Speaker}       
\begin{abstract}
We found solutions to the linear but with complicated kernel and non-homogeneous evolution equations for the cross sections of productions of $n$-cut Balitsky–Fadin–Kuraev–Lipatov (BFKL) Pomerons in the final states of high energy DIS on a nucleus, resumming all multiple rescatterings in the leading logarithmic approximation. For the model leading-twist BFKL kernel, we calculate analytical solutions of these equations by developing the homotopy approach. We also calculate the solution in the large $z=\ln\left(x_{01}^2\,Q_s^2(Y,\mathbf{b})\right)$ and large $n\gtrsim\langle n(z) \rangle$ limits, where $x_{01}$ is the dipole size, $Q_s$ the saturation scale and $\langle n(z) \rangle$ is the average multiplicity of the produced gluons. Having these cross sections we calculate the multiplicity distributions of the produced gluons and describe how the upcoming Electron-Ion Collider (EIC) can test our theoretical formalism.
\end{abstract}




\end{frontmatter}


\section{Introduction}

In recent years, the multiplicity distribution of produced gluons in QCD has attracted considerable attention and has become an active topic of discussion in the high energy QCD community. This interest has been largely motivated by a novel interpretation of the entropy of multiplicity distributions in deep inelastic scattering (DIS). In \cite{KHLE}, it was proposed that the entropy in DIS can be identified with the entanglement entropy between the spatial region probed in the scattering process and the remainder of the proton wave function. The determination of this distribution has traditionally relied on the second approach to the high energy QCD; the Balitsky--Fadin--Kuraev--Lipatov (BFKL) \cite{BFKL,RevLI} Pomeron calculus. In particular, we have the Abramovsky–Gribov–Kancheli (AGK) cutting rules \cite{AGK} for the Pomeron theory of high energy strong interactions, which give us the relation between the total cross section at high energy and multi-particle production processes. In the general case, the AGK formula cannot be summed in closed form and is known only at the level of the initial conditions. However, in DIS in the leading logarithmic approximation (LLA), the BFKL Pomeron calculus is equivalent to the dipole picture of high energy scattering, which is advantageous since the dipole approach allows multiple interactions with the target to be systematically resummed via its evolution equation. In this talk and these proceedings, based mainly on \cite{MDQCD}, we use this equivalence to replace the AGK cutting rules by the evolution equations. In Sect. II we discuss the evolution equations for the cross sections of productions of $n$-cut BFKL Pomerons ($\sigma_n$) at large $N_c$. In Sect. III we study the simplification of these equations for the model leading-twist BFKL kernel and obtain solutions using the homotopy approach and in the large $z\,\&\,n$ limits. We conclude in Sect. IV by summarizing our results.

\section{The cross sections of productions of $n$-cut BFKL Pomerons at high energies}   
In \cite{KLP,Levin:2023mwl}, the small-$x$ evolution equations for the cross sections of productions of $n$-cut BFKL Pomerons ($\sigma_n$) in dipole--nucleus scattering were derived, taking the form: 
\bea \label{ME4}
&&\frac{\partial \,\sigma_n\Lb Y;\vec{x}_{01},\vec{b} \Rb}{\partial \,Y}\,\,=\,\,\frac{\bas}{2 \, \pi} \int d^2 x_2 \,
  \frac{x^2_{01}}{x^2_{02}\,x^2_{12}} \,[ \sigma_n\Lb Y;\vec{x}_{12},\vec{b} \Rb\,\,+\,\,
\sigma_n\Lb Y;\vec{x}_{02},\vec{b}  \Rb\,\,-\,\,\sigma_n\Lb Y;\vec{x}_{01},\vec{b}   \Rb \nn  \\
&&+\,\,\sigma_{n}\Lb Y;\vec{x}_{12},\vec{b}  \Rb  \,\sigma_{SD}\Lb Y;\vec{x}_{02},\vec{b}  \Rb\,\,+\,\,\sigma_{n}\Lb Y;\vec{x}_{02},\vec{b}  \Rb  \,\sigma_{SD}\Lb Y;\vec{x}_{12},\vec{b}  \Rb\nn \,\,+\,\,\sum_{k=1}^{n-1}  \sigma_{n - k}\Lb Y;\vec{x}_{02},\vec{b}  \Rb\,\sigma_{k}\Lb Y;\vec{x}_{12},\vec{b}  \Rb\nn\\ &&-\,\,2\,\sigma_{n}\Lb Y;\vec{x}_{12},\vec{b}  \Rb\,N\Lb Y;\vec{x}_{02},\vec{b}  \Rb\,\,-\,\, 2\,
\sigma_n\Lb Y;\vec{x}_{02},\vec{b}  \Rb
\,N\Lb Y;\vec{x}_{12},\vec{b}  \Rb]
 \eea
Here $Y$ is the total rapidity interval of the collision, $\bas=\as\,N_c/\pi$, boldface variables denote two-component transverse plane vectors with their magnitudes $x \equiv |\vec{x}|$, and differences $x_{ij} \equiv x_i - x_j$. $N\Lb Y;\vec{x}_{ij},\vec{b}  \Rb$ is the imaginary part of the forward dipole--nucleus scattering amplitude for the elastic scattering of the $q\bar{q}$ dipole of transverse size $\vec{x}_{ij}$ at impact parameter $\vec{b}$ obeying the standard Balitsky--Kovchegov (BK) evolution equation \cite{BK} and $\sigma_{SD}\Lb Y;\vec{x}_{ij},\vec{b}  \Rb$ is the single diffractive dissociation cross section with all possible rapidity gaps for the same scattering process obeying the Kovchegov--Levin (KL) evolution equation \cite{KOLE}.

\begin{figure}[h]
 	\begin{center}
 	\leavevmode
 		\includegraphics[width=8cm]{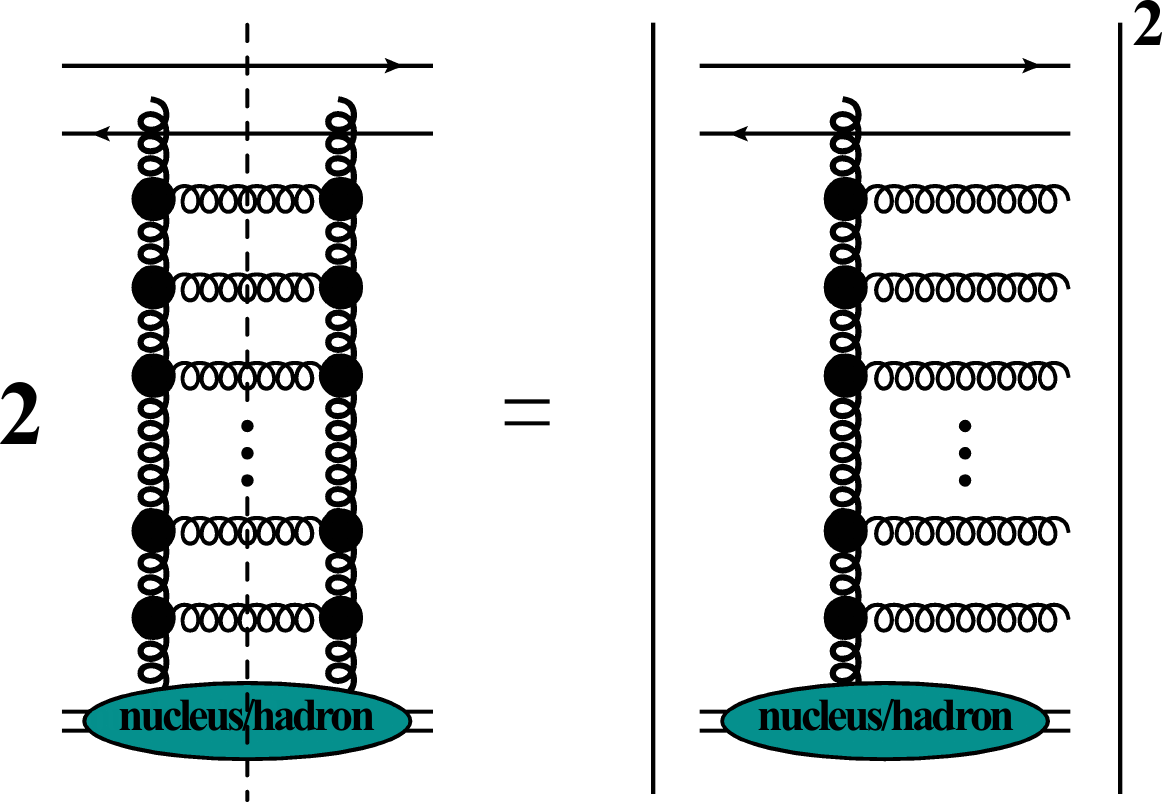}
 	\end{center}
 	\caption{The definition of cut Pomeron through the BFKL
ladder with eﬀective Lipatov vertices (black circles) and reggeized gluons (thick gluon lines). The dashed line shows the cut Pomeron, which
describes the production of gluons.}
 	\label{CutPomeron}
\end{figure}


\begin{figure}[h]
 	\begin{center}
 	\leavevmode
 		\includegraphics[width=12cm]{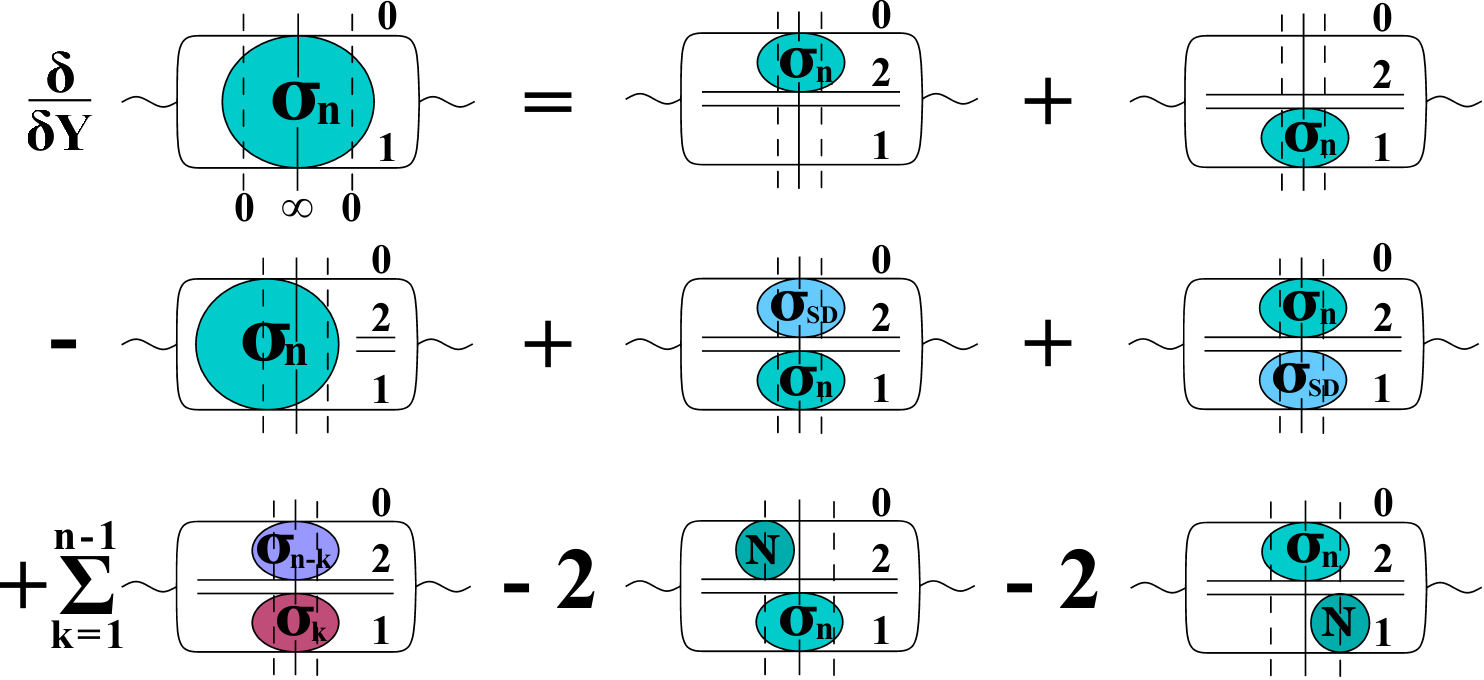}
 	\end{center}
 	\caption{The graphic form of the equation for the cross section of production of $n$-cut BFKL Pomerons. The solid vertical straight line denotes the final state cut, while the dashed vertical straight lines denote the Glauber--Mueller interaction with the target \cite{Mueller:1989st}. All double lines represent gluons at large-$N_c$.}
 	\label{eqn}
\end{figure}

The structure of the cut Pomeron and the evolution equation for the $n$-cut pomeron production is illustrated in \fig{CutPomeron} and \fig{eqn}, respectively. The initial condition for the evolution of $\sigma_n$ is obtained by employing the AGK formula, which gives
\begin{align} 
  \label{AGKIC}
  \sigma_n\Lb Y=Y_A;\vec{x}_{01},\vec{b}\Rb \,\,=\,\ \frac{\Lb\frac{1}{2} \,x_{01}^2 \,
    Q_{s}^2\Lb Y=Y_A,\vec{b}\Rb\Rb^n}{n!}\exp\left\{-\frac{1}{2} x_{01}^2 \,
    Q_{s}^2\Lb Y=Y_A,\vec{b}\Rb\right\}
\end{align}
Here $Q_{s}^2\Lb Y=Y_A,\vec{b}\Rb$ is the initial value of the saturation scale \cite{GLR}, which has the following $Y$ dependence in the fixed coupling case \cite{GLR,MUT,MUPE}:
\beq\label{mtsat}
Q_{s}^2 \Lb Y,\vec{b}\Rb \, = \, Q_{s}^2\Lb Y=Y_A,\vec{b}\Rb \, e^{\bas\,\kappa \,Y -\,\,\frac{3}{2\,(1-\gamma_{cr})} \ln \left[\bas\,Y \right]}
\eeq
where $\kappa$ and $\gamma_{cr}$ are determined by the following equations:
\beq \label{GACR}
\kappa \,\,\equiv\,\, \frac{\chi\Lb \gamma_{cr}\Rb}{1 - \gamma_{cr}}\,\,=\,\, - \frac{d \chi\Lb \gamma_{cr}\Rb}{d \gamma_{cr}}~~~\,\,\,\mbox{and}\,\,\,~~~\chi\Lb \gamma\Rb\,=\,\,2\,\psi\Lb 1 \Rb\,-\,\psi\Lb \gamma\Rb\,-\,\psi\Lb 1 - \gamma\Rb
\eeq
$\chi(\gamma)$ is the eigenvalue of the BFKL kernel \cite{BFKL} and $\psi(\gamma)=d\ln\Gamma(\gamma)/d\gamma$ is the Euler psi-function. 
\section{The solution for the simplified BFKL kernel}
In the kinematic regions: $ x_{01} \approx x_{02} \gg x_{12}\gg 1/Q_s$  and $ x_{01} \approx x_{12} \gg x_{02}\gg 1/Q_s$ \cite{LETU}, \eq{ME4} reduces to:
\bea \label{EQXS01}
   &&\frac{\partial\,\sigma_n\left( Y,\mathbf{x}_{01},\mathbf{b}\right)}{\partial Y} \,\,=\,\,\bar\alpha_S\Bigg\{-\,\ln\left( x_{01}^2\,Q_s^2(Y,\mathbf{b})\right)\,\sigma_n\left( Y,\mathbf{x}_{01},\mathbf{b}\right) \,\,+\,\,\sigma_n\left( Y,\mathbf{x}_{01},\mathbf{b}\right)\,\int\limits_{Q_{s}^{-2}(Y,\mathbf{b})}^{x_{01}^2} \frac{dx_{02}^2}{x_{02}^2}\,\Delta\left( Y,\mathbf{x}_{02},\mathbf{b}\right)  \nn\\&&+\,\,\Delta\left( Y,\mathbf{x}_{01},\mathbf{b}\right)\,\int\limits_{Q_{s}^{-2}(Y,\mathbf{b})}^{x_{01}^2}\frac{dx_{02}^2}{x_{02}^2}\,\sigma_n\left(Y,\mathbf{x}_{02},\mathbf{b}\right)\,\,+\,\,\sum_{k=1}^{n-1} \,\int\limits_{Q_{s}^{-2}(Y,\mathbf{b})}^{x_{01}^2}\frac{dx_{02}^2}{x_{02}^2}\,\sigma_{n - k}\left( Y,\mathbf{x}_{02},\mathbf{b}\right)\,\sigma_{k}\left( Y,\mathbf{x}_{01},\mathbf{b}\right)\Bigg\}
\eea
where $\Delta\,=\,1\,-\,2\,N\,+\,\sigma_{SD}$. Defining the geometric scaling variables $z \,=\, \ln\left[  x^2_{01}\,Q^2_s\Lb Y,\vec{b}\Rb\right]$ $\Lb z'  \,=\,\ln\left[  x^2_{02}\,Q^2_s\Lb Y,\vec{b}\Rb\right]\Rb$ and the rescaled rapidity variable $\delta\tilde Y\,=\,\bar\alpha_S\left( Y\,-\,Y_A\right)$, the equations become:
\bea \label{EQXS02}
   &&\frac{\partial\,\sigma_n\Lb \dY,z\Rb}{\partial \dY}\,\,+\,\,\kappa\,\frac{\partial\,\sigma_n\Lb \dY,z\Rb}{\partial z}\,\,=\,\,-\,z\,\sigma_n\Lb \dY,z\Rb\nn\\&&+\,\,\sigma_n\Lb \dY,z\Rb\,\intl_{0}^{z}\,dz'\,\Delta\Lb \dY,z'\Rb \,\,+\,\,\Delta\Lb \dY,z\Rb\,\intl_{0}^{z}dz'\,\sigma_n\Lb \dY,z'\Rb\,\,+\,\,\sum_{k=1}^{n-1}  \,\intl_{0}^{z}dz' \,\sigma_{n - k}\Lb \dY,z'\Rb\,\sigma_{k}\Lb \dY,z\Rb
\eea
Assuming that we have geometric scaling symmetry \cite{GS}, which is valid deep inside the saturation region, we simplify \eq{EQXS02} to
\beq \label{MDLN}
\kappa \,\frac{ d \,\sigma_{n}\Lb z \Rb}{d\,z}\,\,=\,\,-\,z\,\sigma_{n}\Lb z\Rb \,\,+\,\, \Delta\Lb z \Rb \,\Sigma_{n}\Lb z\Rb \,\,+\,\, \Sigma_\Delta\Lb z \Rb\,\sigma_{n}\Lb z\Rb\,\,+\,\,\sum_{k=1}^{n-1}\Sigma_{n-k}\Lb z \Rb \,\sigma_{k}\Lb z \Rb
\eeq
where $\Sigma_\Delta\Lb z \Rb = \intl^z _{0} d z' \Delta\Lb z'\Rb$ and $\Sigma_{n}\Lb z\Rb = \intl^z _{0}d z' \sigma_{n}\Lb z'\Rb$. In the next subsections, we present two approaches to solving \eq{MDLN}: i) the homotopy approach and ii) the large $z\,\&\,n$ approach.
\subsection{The homotopy solutions}
This method we have discussed in our previous papers (see Refs.\cite{HomApproach}). It is based on the idea \cite{HE1} that we can divide the general nonlinear equation in two parts:
 \beq \label{HOM1}
\mathscr{L}[\sigma_n] +  \mathscr{N_{L}}[\sigma_n]=0
\eeq 
 where $\mathscr{L}[\sigma_n] $ include the linear evolution and part of the nonlinear corrections, which can be treated analytically (or almost analytically).
 The non-linear part $
 \mathscr{N_{L}}[\sigma_n] $ has an arbitrary form. As a solution, we introduce  the following  equation for the homotopy function $ {\mathscr H}\Lb p,\sigma_n\Rb$:
 \beq \label{HOM2}
 {\mathscr H}\Lb p,\sigma_n\Rb\,\,=\,\,\mathscr{L}[\sigma_n^{\Lb p\Rb}] \,+ \,  p\, \mathscr{N_{L}}[\sigma_n^{\Lb p\Rb}] \,\,=\,\,0
 \eeq
 
 Solving \eq{HOM2} we reconstruct the function
 \beq \label{HOM3}
 \sigma_n^{\Lb p\Rb}\Lb Y;  \x_{01},  \vb\Rb\,\,=\,\, \sigma_n^{\Lb 0\Rb}\Lb Y;  \x_{01},  \vb\Rb\,\,+\,\,p\, \sigma_n^{\Lb 1\Rb}\Lb Y;  \x_{01},  \vb\Rb \,+\,p^2\, \sigma_n^{\Lb 2\Rb}\Lb Y;  \x_{01},  \vb\Rb \,\,+\,\,\dots
 \eeq
 with $\mathscr{L}[\sigma_n^{\Lb 0\Rb}] = 0$. \eq{HOM3}  gives  the solution to the non-linear equation at $ p = 1$.  The hope is that several  terms in series of \eq{HOM3} will give a good  approximation in the solution of the equation. We take:
  \begin{subequations} 
\bea
\mathcal{L}[ \sigma_n]\,\, &=&\,\, 
\kappa\,\frac{d\sigma_n\Lb z\Rb}{dz} \,\,+\,\, \Lb z\,-\,\Sigma_\Delta\Lb z\Rb\Rb\,\sigma_n\Lb z\Rb \,\,-\,\,\sigma_{0,n}\,\Delta\Lb z\Rb\,\,-\,\,\sum_{k=1}^{n-1}\Sigma_{n-k}\Lb z \Rb \,\sigma_{k}\Lb z \Rb\label{EQXS11a}\\
\mathcal{N}_{\mathcal{L}}[ \sigma_n]\,\,&= &\,\,\Delta\Lb z\Rb\,\tilde\Sigma_n\Lb z\Rb \label{EQXS11b}\eea
\end{subequations}
where $\sigma_{0,n}=\intl_{0}^{\infty} dz' \,\sigma_n(z')$ is a constant and $\tilde\Sigma_n(z)=\intl_{z}^{\infty} dz' \,\sigma_n(z')$. Deep inside the saturation regime ($z\gg1$), the zeroth iteration gives
\beq\label{XSN3}
\sigma^{(0)}_{n}\Lb z\Rb\,\,=\,\,\exp\left\{ - \frac{ z^2}{2 \,\kappa }\right\} \left[ C^{\Lb n\Rb}\,\,+\,\,\frac{1}{\kappa}\int_0^z dz'\exp\left\{ \frac{ z'^2}{2 \,\kappa }\right\}\,\left(\sigma_{0,n}\,\Delta\Lb z'\Rb\,\,+\,\,\sum_{k=1}^{n-1}\Sigma_{n-k}\Lb z' \Rb \,\sigma_{k}\Lb z' \Rb\right)\right]
\eeq
where $C^{(n)}$ is a constant to be fixed from the initial conditions. Calculating the next iterations in this approach, we find that the ratios are indeed small, and decrease as the number of iterations increases (see \cite{MDQCD} for more details). However, this approach is valid only for small $n$, as shown in \cite{MDQCD}. For large $n$, we discuss the solution in the next subsection.
\subsection{The large z and $n\,\gtrsim\,\langle n(z) \rangle$ solutions}
Suggesting a solution of the form:
\beq \label{MDLN1}
\Sigma_n\Lb z\Rb\,\,=\,\,\phi\Lb z\Rb \,\exp\left\{ - n \,\Phi\Lb z \Rb\right\}
\eeq
with $\phi$ and $\Phi$ unknown functions, we substitute \eq{MDLN1} into \eq{MDLN} and obtain:
\beq
\kappa\,\frac{d^2\Sigma_n}{dz^2}\,\,=\,\,-\frac{d}{dz}((z\,-\,\Sigma_\Delta(z))\,\Sigma_n(z))\,\,+\,\,\Sigma_n(z)\,\,+\,\,\frac{n-1}{2}\,\frac{d}{dz}(\phi(z)\,\Sigma_n(z))
\eeq
Introducing $\mathcal{S}_n\left( z\right) =\int\limits^z d z' \,\Sigma_n\left( z'\right)$ and assuming that all functions decreases at large $z$, we have 
\beq
\kappa \,\frac{d \Sigma_n}{d z}\,\,=\,\,-   (z\,-\,\Sigma_\Delta(z)) \,\Sigma_n( z) \,\,+\,\, \mathcal{S}_n( z)\,\,+\,\, \frac{n-1}{2} \,\phi\left( z\right)\,\Sigma_n\left( z\right)
\eeq
The resulting equations can be solved and all unknown function can be found. Taking the large $z\,\&\,n$ limits, we find for DIS:
\beq \label{MDLN15}  
          \sigma_n\Lb z \Rb\,\,=\,\,\frac{d}{dz}\Sigma_n(z)\,\,\xrightarrow{n \geq \langle n(z) \rangle}\,\, \frac{2}{ \kappa}\,\frac{z^2}{\langle n(z) \rangle}\,\Psi\Lb\frac{n}{\langle n(z) \rangle}\Rb 
\eeq
where $\Psi\Lb \xi\Rb= \xi\,e^{- \xi}$ is a universal scaling function known as the Koba--Nielsen--Olesen (KNO) scaling function \cite{KNOscal} and $\langle n(z) \rangle$ is the average multiplicity distribution which takes the form:
    \beq \label{MDLN170}    
      \langle n(z) \rangle \,\,=\,\,2\,n_0\,z\,\exp\left( \frac{z^2}{2\,\kappa}\right)
    \eeq
where $n_0$ is a constant to be determined from the matching of \eq{MDLN15} with the initial condition of \eq{AGKIC}.
We now calculate the multiplicity distribution for the produced gluons, obtaining:
   \beq\label{MDLN17}
P_n\Lb z\Rb\,\,\equiv\,\,\frac{\sigma_n\Lb z\Rb}{\sigma_{in}\Lb z\Rb}\,\,=\,\,\frac{2}{ \kappa}\,\frac{z^2}{\langle n(z) \rangle} \,\Psi\Lb \frac{n}{\langle n(z) \rangle}\Rb 
\eeq 
where we used the unitarity limit of $\sigma_{in}(z)=\sum_{n=1}^\infty \sigma_n\Lb z\Rb\rightarrow 1$ at large $z$. Finally, the cross sections for $k$-cut Pomerons need to be convoluted with the Poisson distribution of the produced gluons in the pomeron with average number of gluons $k\,\Delta_{\mbox{\tiny BFKL}}\,Y$ to obtain the produced gluons, namely
 \beq \label{I2}
\tilde{ \sigma}_n\Lb Y;\vec{x}_{01},\vec{b} \Rb\,\,=\,\,\sum_{k=1}^{\infty}\underbrace{ \sigma_k\Lb Y;\vec{x}_{01},\vec{b} \Rb}_{ \propto\,\Lb\sigma^{\mbox{\tiny BFKL}}_{in}\Lb Y;\vec{x}_{01},\vec{b} \Rb
 \Rb^k}\,\underbrace{\frac{\left(k\,\Delta_{\mbox{\tiny BFKL}}\,Y\right)^n}{n!}\,e^{-k\,\Delta_{\mbox{\tiny BFKL}}\,Y}}_{\mbox{Poisson distribution}}\,\,\xrightarrow{Y \,\gg\,1} \,\, \sigma_{k = n/(\Delta_{\mbox{\tiny BFKL}}\,Y)}\Lb Y;\vec{x}_{01},\vec{b} \Rb
\eeq
where $\Delta_{\mbox{\tiny BFKL}}$ is the BFKL Pomeron intercept. Let us note also that the “produced” gluons are understood to be eventually convoluted with fragmentation functions, according to the standard perturbative QCD prescription.
\section{Conclusions}\label{sec:sumandconc}
In this paper, we have solved the evolution equations for the cross sections of productions of $n$-cut BFKL Pomerons in high energy DIS on a nucleus, using two different approaches: the homotopy approach, and the large $z$ and large $n\gtrsim\langle n(z) \rangle$ limits. The solution at large $n$ has an $n$ dependence different from that of the homotopy solution. A matching between the two solutions is therefore not possible. However, it turns out that the main contribution to the inelastic cross section comes from large $n$ solution. To provide a matching is a big problem for the future.
We also calculate the multiplicity distribution $P_n(z)$ for the large $z$ and large $n$ cases. For small $k$, one can instead compute the $k$-th moments of the multiplicity distribution using the relation $\langle n^k(z)\rangle = \sigma_k(z)\,n_{\mbox{\tiny BFKL}}^k(z)$, where $n_{\mbox{\tiny BFKL}}(z)$ denotes the multiplicity of produced gluons in the BFKL Pomeron \cite{AGK,MUMULT}. Having these moments, one can then compute the corresponding cumulants. We plan to calculate these quantities in future publications, to compare them with experimental data from HERA \cite{HERAH1}, and to make predictions for the future Electron-Ion Collider (EIC) \cite{EIC}.

\section*{Acknowledgments}
We are very grateful to E. Levin and Y. Kovchegov for helpful discussions. This research was supported by Fondecyt (Chile) grant No. 1231829. J. G. acknowledges support of the fellowship ``Beca de T\'ermino de Tesis de Doctorado $N^o$ 079/2025'' of Direcci\'on de Postgrado (DP), UTFSM, and expresses his gratitude to the Institute of Physics of PUCV (IFIS). J.G. also thanks the organizers of Baryons 2025 for support.

\end{document}